# Room-scale magnetoquasistatic wireless power transfer using a cavity-based multimode resonator

Takuya Sasatani[1]*     (ORCID ID: 0000-0003-2268-6106)

Alanson P. Sample[2]    (ORCID ID: 0000-0002-8046-0538)

Yoshihiro Kawahara[1]   (ORCID ID: 0000-0002-0310-2577)

(*Corresponding author) Email: sasatani@akg.t.u-tokyo.ac.jp

[1] Department of Electrical Engineering and Information Systems, Graduate School of Engineering, The University of Tokyo, Bunkyo-ku, Tokyo, Japan

[2] Electrical Engineering and Computer Science Department, University of Michigan, Ann Arbor, Michigan, USA

**Abstract**

Magnetoquasistatic wireless power transfer can be used to charge and power electronic devices such as smartphones and small home appliances. However, existing coil-based transmitters, which are composed of wire conductors, have a limited range. Here, we show that multimode-quasistatic cavity resonance can provide room-scale wireless power transfer. The approach uses multidirectional, widely distributed currents on conductive surfaces that are placed around the target volume. It generates multiple, mutually unique, three-dimensional magnetic field patterns, where each pattern is attributed to different eigenmodes of a single room scale resonator. Using these modes together, a power delivery efficiency exceeding 37.1% can be achieved throughout a 3 m × 3 m × 2 m test room. With this approach, power exceeding 50 W could potentially be delivered to mobile receivers, while in accordance with safety guidelines.

Electronic devices, including cell phones, sensor nodes, and robots, are widely used in industrial and personal living spaces. However, the devices are primarily powered by wired connections or disposable batteries, which require manual operation and may adversely affect the environment. The benefits of untethering these devices are widely acknowledged and are highlighted by the recent commercialisation of wireless communication technologies[1-3]. Wireless power transfer technologies, which are capable of safe power transfer to numerous devices scattered over large three-dimensional volumes, could potentially be used to untether the devices. However, the trade-off between deliverable power levels and transfer distance has hindered the establishment of such technologies.

Early attempts at wireless power transfer were mostly based on electromagnetic (EM) radiation (*e.g.* microwaves)[4–6]. Although modern beamforming techniques now enable efficient power delivery over a certain distance, they require large antenna arrays and complicated mechanisms for continuous tracking. Moreover, safety concerns may arise because microwave causes relatively large interaction with biological tissue[7,8]. To transmit higher power levels without detrimental health effects, magnetic resonance coupling, which transmits power between a pair of coil-based resonators using magnetic fields, has been explored[9–18]. However, the rapid decrease in field intensity with increasing distances limits the power transfer range of this approach to approximately the diameter of the coils. Additionally, asymmetric systems, with a large transmitter and a small receiver, exhibit low efficiency due to insufficient coupling, and using multiple coils only extends this range to a two-dimensional surface.

Quasistatic cavity resonance (QSCR) uses a room scale cavity with a central pole and inserted lumped capacitors[19,20]. This approach generates a 3D magnetic field distribution within the room via the current widely distributed on the structure. Compared to using conventional cavity resonators for wireless power[21,22], this technique has the advantage of confining the electric field, which mainly interferes with biological tissues, within the lumped capacitors, while also providing an ability to tune the system resonant frequency by adjusting the values of the lumped capacitors. However, the field distribution of QSCR is non-uniform, and consequently there are large portions of the room volume where power delivery is inefficient.

Recent analysis has indicated that efficient power transfer over a larger volume could be achieved by extending QSCR to multimode structures[23]. In this Article, we experimentally show that this can be achieved using a room-scale multimode resonator that generates two unique and widely distributed magnetic field patterns, which cover the entire volume when used together. Using these field patterns, power can be efficiently delivered throughout a test room with dimensions of 3 m × 3 m × 2 m. The approach — termed multimode quasistatic cavity resonance (M-QSCR) — uses a resonant structure composed of conductive surfaces and lumped capacitors to accommodate multiple resonant modes with mutually unique patterns of the oscillating surface current. With the approach, we show that electronic devices such as stand lights, fans, and smartphones can be wirelessly powered in a furnished room. Simulations also show that our room-scale resonator can safely deliver over 50 W of power to devices as small as 1/5000 of the powering range.

## Multimode room-scale resonator

Coil-based methods have suffered from the narrow powering range (Fig. 1(**a**)) and previous QSCR-based methods have only supported a single current and magnetic field distribution, occurring areas where power delivery is inefficient (Fig. 1(**b**)). Our multimode approach (Fig. 1(**c**)) can selectively generate multiple magnetic field patterns based on the input tone of the external drive coil, which is inductively coupled with the room-scale transmitter resonator. Thereafter, the receivers positioned within the range of the generated magnetic field are powered (Fig. 1(**d**)). As the generated field patterns complement the null zones of each other, the entire volume can be covered efficiently. This cannot be achieved by using either mode alone. Furthermore, by using one of the existing resonant modes, the obstructive central conductive pole, which was necessary for single-mode QSCR, can be omitted, while still maintaining a large power delivery range. (M-QSCR is compared to previous approaches in Supplementary Table 1.)

Surface conductors such as aluminium sheets naturally support current distributions along various directions. However, methods of accommodating multiple high-quality (Q) factor resonant modes, where each mode generates a magnetic field pattern that covers mutually different large portions of the volume, remain unclear. As the proposed approach is based on magnetoquasistatic fields, resonance can be understood as the condition under which magnetic energy, possessed by the widely distributed current, and electrical energy, confined in the lumped capacitors, are balanced. Therefore, the arrangement of the conductive surfaces that generate the widely distributed magnetic field patterns and the effective insertion of the lumped capacitors within the assumed current loops are primary criteria for designing the structure.

Accordingly, we designed a room-scale resonant structure that accommodates the two widely distributed current patterns shown in Fig. 1(**c**). One current pattern is mainly distributed near the centre of the room (*i.e.* at the pole), and the current directions on all walls are identical, serving as the return path for the intense pole current. Meanwhile, the other current pattern is distributed near the edges (*i.e.* the walls of the structure), and the current directions on adjacent walls are opposite, serving as the return path of the adjacent walls' current. We assumed that these distributions would compose unique magnetic field patterns that cover the null zones of each other (see Methods and Supplementary Fig. 1 for details). An overview of the designed room-scale resonator that enables these current flows and a photograph of the constructed wireless power transfer system are shown in Fig. 2(**a**) and (**b**). The covered charging volume is shown in Supplementary Fig. 2, and the mounted lumped capacitors are shown in Supplementary Fig. 3 (see Methods for details regarding the covered area and its construction). This room-scale resonant structure is composed of (i) an aluminium box with the corners cut off and lumped capacitors inserted at the centre of each wall, and (ii) a conductive pole with inserted lumped capacitors. This structure accommodates two resonant modes that possess unique current distribution and magnetic field patterns, as shown in Fig. 2(**c**)–(**f**).

We termed these two modes as the pole independent (PI) mode and the pole dependent (PD) mode based on the following observations. When the PI mode is excited, current only flows through the walls, ceiling, and floor and generates an intense magnetic field distribution near the walls. Conversely, the PD mode induces a current that flows through the central pole, which generates a magnetic field that circulates around the pole and is most intense near the central pole. These two modes yield magnetic field distributions that cover the weak areas of each mode. Therefore, by multiplexing these resonant modes, an intense magnetic field permeating a larger volume than that covered by either mode alone can be achieved; thus, high-efficiency power delivery throughout the volume can be realised. For the PI mode, the current and magnetic field intensity decrease to zero near the centre of the room. Therefore, the presence of the central pole does not affect this mode. Consequently, wireless power transfer over a large portion of the volume remains enabled even when the central conductive pole is omitted. Inner views of the room-scale resonator configured with and without the central pole are provided in Supplementary Fig. 4(**a**) and (**b**). An overview of the room-scale resonator, including the resonator structure, mounted lumped capacitors, covered range, and an animation of the oscillating current/magnetic field, is provided in Supplementary video 1.

## Analyses and tuning of resonant frequencies

Although the resonant frequency of typical $LC$ resonators can be tuned using lumped capacitors, the widely used expression for the resonant frequency of an $LC$ tank is not directly applicable to the resonator in this study. For the proposed resonator, it is difficult to define the inductance $L$ because the current is widely distributed over the surfaces without converging to a single port and four independent current loops exist within each current distribution (Supplementary Fig. 5 and Supplementary Fig. 6). Furthermore, relevant regulations regarding the safety of wireless power transfer restrict the operating frequencies of such systems. Therefore, the following analysis is performed for enabling the independent tuning of each resonant frequency of the structure to predefined values.

For tuning the resonant frequency of each mode, $f_{\text{PI}}$ and $f_{\text{PD}}$, the structure is assumed to have an equivalent inductance $L_{\text{eq}}$. Then, the angular resonant frequency is determined using the conventional expression for an $LC$ tank, *i.e.* $2\pi f_0 = \omega_0 = 1/\sqrt{LC}$, whereas the capacitance $C$ is set by using discrete capacitors with known values. Analytical expressions for the remaining parameter, equivalent inductance $L_{\text{eq}}$, are difficult to derive, as mentioned above. Hence, to subvert manual calculation, we developed an approach based on the finite element method (FEM) for computing the inductance of a structure with a fixed geometry.

Under both modes, the surface current, $J_S$, does not cross the $J_S = 0$ boundaries shown in Supplementary Fig. 5(**a**) and (**b**). Therefore, the total current can be analysed as a combination of four independent current loops flowing through the unit cells shown in Supplementary Fig. 6, which are coupled with each other. Thus, owing to the symmetry of the structure, the magnetic energy corresponding to a unit current loop, $w_{\text{unit}}$, should be equal to a quarter of the total energy in the entire volume. If the eigenmodes are solved using FEM simulations involving known capacitance values and a fixed structure, the software can then be used to evaluate the current in a unit loop, $I_{\text{unit}}$, and the total magnetic energy of the structure, $\alpha$. On dividing $\alpha$ by four, we obtain $w_{\text{unit}}$.

Finally, we determined the equivalent inductance by noting that the energy stored in an inductor (a unit cell) is $\alpha/4 = w_{\text{unit}} = L_{\text{eq}} I_{\text{unit}}^2/2$, where $\alpha$ and $I_{\text{unit}}$ are obtained from the simulations. After computing $L_{\text{eq}}$, the desired resonant frequency can be determined from the following expression:

$$\omega_0 = \frac{1}{\sqrt{L_{\text{eq}} C}} = \frac{1}{\sqrt{\frac{\alpha}{2I_{\text{unit}}^2} \cdot C}} \tag{1}$$

Considering the capacitor positions and current paths shown in Fig. 2(**a**), (**c**), and (**e**), the capacitance inserted in the unit current loops of the PD and PI modes can be calculated as the following by applying fundamental circuit analysis:

$$C_{\text{PD}} = \frac{C_{\text{pole}} C_{\text{wall}}}{C_{\text{pole}} + 4 C_{\text{wall}}} \tag{2}$$

$$C_{\text{PI}} = \frac{C_{\text{wall}}}{4} \tag{3}$$

Here, $C_{\text{pole}}$ and $C_{\text{wall}}$ are the lumped capacitor values inserted in the gap of the pole and each wall. Using these expressions together with Equation (1), the two resonant frequencies can be tuned independently. The resonant frequencies of the PD mode, $f_{\text{PD}}$, and the PI mode, $f_{\text{PI}}$, are set to 1.20 MHz and 1.34 MHz, respectively, through the process described above. The quality factors of each mode, measured via an established method[24], were 1230 (PD mode) and 615 (PI mode).

**Power transfer efficiency**

Next, we determine the power transfer efficiency between the 3 m × 3 m × 2 m room-scale resonator and the 150 mm × 150 mm receiver through simulations and measurements. The receiver is as small as 1/5000 of the powering range (see Methods for details regarding the range and scaling); this can be considered as a highly asymmetric configuration, which typically suffer from insufficient coupling. The power transfer efficiency throughout the volume was determined based on the coupled mode theory by using the measured Q factors and the simulated coupling coefficients[19,21,25] (see Methods for details). In addition, to confirm these results, efficiency measurements based on the measured S-parameters were conducted at the extracted positions[26] (see Methods for details). For the receiver, we used a six-turn, 150 mm × 150 mm square coil with a series capacitor connected to tune the resonance frequency, as shown in Supplementary Fig. 7(**a**). The Q-factor of this receiver was 236. In the measurements, an external drive coil, shown in Supplementary Fig. 7(**b**), was used to stimulate the resonant modes of the room-scale resonator. In both the coupled mode theory-based evaluations and the S-parameter-based evaluations, the value of load impedance is assumed to be the one that maximises the power transfer efficiency; this is a reasonable and typical assumption considering the development of maximum efficiency point tracking methods[18, 26].

Fig. 3(**a**)–(**e**) show the power transfer efficiency along the $z = 0$ and $y = 0$ planes evaluated based on the coupled mode theory. These results show that the system forms a 3D powering range, and the two modes successfully complement the null zone of each other. The evaluations throughout the entire volume revealed that the proposed approach achieved an efficiency exceeding 50% in 98.0% of the room volume by multiplexing the PI and PD modes based on the receiver position. Moreover, the power transfer efficiency was greater than 37.1% at any position within the room volume. This analysis also revealed that the PD mode alone, which closely resembles the previous single-mode QSCR, can achieve an efficiency exceeding 50% in only 57.5% of the room volume, and its efficiency decreases to 1% near the walls. These results show that the proposed approach covers the entire volume of the room, while the PD mode alone (resembling the previous method) only covers approximately half of the room (considering positions where the efficiency exceeds 50%). Additionally, when only the PI mode is used (*i.e.* when the central conductive pole is omitted), an efficiency exceeding 50% could be obtained in 52.4% of the room volume. Overall, these results indicate the following: (i) by leveraging multiple resonant modes (*i.e.* magnetic field patterns), a significantly larger portion of the volume could be covered, as compared to that covered by either mode alone; (ii) the power transfer range successively forms a 3D volume, and the transfer

efficiency has a limited dependence on z-coordinates; and (iii) the resonant modes' repertory allows for the omission of the central pole, which was essential for previous approaches, while still covering a large portion of the room.

Lastly, we measured the power transfer efficiency at extracted positions within the room volume to confirm the results based on the coupled mode theory; this measurement was based on the S-parameters obtained by two-port measurements using a vector network analyser (VNA). Line (i) and line (ii) in Fig. 3(**f**) depict the positions where the receiver was located for this measurement. As shown in Fig. 3(**f**), the drive coil was placed such that it efficiently coupled with the magnetic field patterns of the corresponding resonant mode. Fig. 3(**g**) and (**h**) present a comparison of the power transfer efficiency determined using the measured S-parameters and that determined based on the coupled mode theory. The validity of the coupled mode theory-based analysis conducted across the entire room volume was proved by the good agreement between these results.

We note that the receiver orientation and the coil size would also affect the power transfer efficiency. We have discussed this issue in the Conclusion section and in Supplementary Notes 1 and 2.

**Safety**

For deploying wireless power transfer systems in public environments, it is necessary to evaluate the effect of the exposed EM field on people present within the charging coverage[27]. As measures for safety, the Federal Communication Commission (FCC) and the Institute of Electrical and Electronics Engineers (IEEE) have established guidelines pertaining to EM field exposure. The primary health effects caused by EM field exposure in the used frequency ranges are temperature elevation and electrostimulation. Therefore, relevant standards impose basic restrictions based on the power absorbed by biological tissues (*i.e.* specific absorption rate: SAR) and the internal electrical field, which are the physical quantities associated with these two health effects[28].

Considering IEEE standards, previous research suggests that the SAR is a more restrictive factor than the internal electric field in the frequency bands used in our system[28–30]; thus, we analyse the localised heat stress and full-body heating through the following analysis[27,31]. This analysis considers the compliance with restrictions on uncontrolled exposure to the general public, which limits the average SAR of the entire body to 0.08 W/kg and the localised SAR, defined as the average power absorbed by a 1 g tissue sample, to 1.6 W/kg (see Methods for details)[30].

The SAR is dependent on the input power and the power transfer efficiency[19]. For instance, if the system is driven under non-loaded (*i.e.* zero-efficiency) conditions, exposure levels tend to reach the regulatory limits even with a low input power, as compared to high-efficiency conditions. This is because an increase in the power transfer efficiency increases the proportion of power delivered to the receiver; consequently, the proportion of reactive energy stored in the room-scale resonator decreases. As discussed in the previous section, the proposed system achieves a power transfer efficiency exceeding 50% in most (over 98%) of the room volume. Therefore, for subsequent analyses regarding safety, we investigated the input power limit at which the SAR reaches regulatory thresholds when the system operates at a power transfer efficiency of 50%.

We conducted simulations of SAR using Hyperworks FEKO, which is an EM solver that supports the method of moments and FEM hybrid simulations; surroundings of the human body model were analysed via FEM, which can calculate the SAR without a loss in accuracy within domains considerably smaller than the wavelength[30]. For the simulations, a 1.88 m full-body anatomical model comprising each tissue component annotated with its EM properties and mass was employed. This model was placed on three positions along the line between the centre of the room and its walls (denoted by $d$ in Fig. 4(**a**)). Two resonant modes were considered for each placement of the body model; thus, a total of six configurations were analysed.

The obtained maximum input power values for the abovementioned configurations are shown in Fig. 4(**b**). For every configuration, the threshold of localised SAR was reached earlier than the threshold of the whole-body average SAR. As these results are based on a power transfer efficiency of 50%, the power delivered to the load is half of the input power. The input power threshold is greater than 100 W for all the tested configurations, and the results suggest that the input power can potentially be increased further if people are appropriately positioned or restricted from certain areas. To provide a graphical representation of heat stress, the distribution of the SAR at the threshold of input power is shown in Fig. 4(**c**). Here, 0 dB corresponds to the threshold for localised SAR (1.6 W/kg), whereas -13 dB corresponds to the threshold for average SAR (0.08 W/kg).

**Wireless power transfer in a furnished room**

To demonstrate the applicability of the proposed approach in daily life, the constructed wireless power room was furnished with typical devices such as floor lamps and smartphones. Fig. 5 shows a photograph of the fully functional wireless power transfer room, where the central conductive pole is omitted and only the PI mode is used (Supplementary video 2). The labelled devices were augmented with wireless power transfer receivers (see Supplementary Fig. 8 for details). As shown in Fig. 5(**a**), the proposed technology enables the seamless untethering of electronic devices without interfering with conventional lifestyles.

**Conclusion**

We have shown that safe wireless power transfer can be achieved within 3D volumes using M-QSCR. Analyses and measurements show that, by leveraging the multidirectional and widely distributed surface currents, efficient power transfer can be achieved throughout a room. In addition, by using the range of available resonant modes, the central pole, which was essential in the previous single-mode approach, can be omitted. As our approach uses magnetoquasistatic fields, room dimensions and receiver sizes can be flexibly determined, independent of the operating wavelength, by adjusting the values of the lumped capacitors. Another benefit of using magnetoquasistatic fields is that the existence of everyday objects does not notably affect the system operation. Given the flexibility in system dimensions and the robustness to interference caused by external objects, we expect that our technology should be widely applicable in a range of scenarios, including charging cabinets, wireless charging rooms, and untethered factories. One limitation is that the receiver coil needs to be orientated orthogonally with the ambient magnetic field to obtain the maximum efficiency at each point (see Supplementary Note 1 and Supplementary Figs. 9-10 for details). However, because our work extends the powering range to full-volume, several peripheral technologies could overcome this challenge in the future[32,33] (see Supplementary Note 2 and Supplementary Fig. 11 for details).

# Methods

### Cutting the corners of the room

To design two current directions distributed near the centre (*i.e.* at the pole) and near the walls of the room, we focused on the direction of the wall current flowing between the floor and ceiling. For generating intense current at the central pole (PD mode), the wall current should be in the opposite direction of the pole current, collectively acting as a "return path" for the intense central pole current. Supplementary Fig. 1 (**a**) shows the top view of such a current flow. Meanwhile, for focusing the current near the structure's walls (PI mode), the currents should form a closed loop within the walls. An approach for designing such a current path is to make the wall current's direction of adjacent walls opposite, making the adjacent wall work as the return path of each other. Supplementary Fig. 1 (**b**) shows the top view of such a current flow. We assumed that this approach considering the current direction of adjacent walls and accommodating the current flowing intensely in different areas of the structure, would compose unique magnetic field patterns that cover each other's null zones.

Next, we evaluated the effect of the size of the wall opening on the magnetic field distribution. Supplementary Fig. 1 (**c-f**) shows the field distribution when the wall width proportion $p$ (see Supplementary Fig. 1 (**a**)) is varied as 0.3, 0.5, and 0.7. As shown in Supplementary Fig. 1 (**c-f**), the wall size slightly alters the magnetic field distribution within the cavity, although the primary trend is invariant. To investigate a basic configuration, we set the wall width to half of the covered area's width ($p = 0.5$) for the experiments and in-depth analyses.

### Construction of multimode QSCR and the charging area

The dimensions of the constructed M-QSCR test room were 3 m × 3 m × 2 m. Supplementary Fig. 2 shows the charging area in this room. The room dimensions were determined based on the largest space we could find. Although the resonant structure has an octagonal shape, the charging volume is rectangular. Therefore, a rectangular room can be created using walls, a floor, and a ceiling within the charging area, as shown in Supplementary Fig. 2 and Supplementary video 1; the images in Fig.

5 were obtained in this manner. The size ratio between the receiver size (Supplementary Fig. 7(**a**), 150 mm × 150 mm square coil) and the charging area was defined as follows:

$$(\text{Length of one side of the receiver})^3 : \text{room volume}$$

$$= 0.15^3 \text{ m}^3 : 18 \text{ m}^3$$

$$\sim 1 : 5000 \qquad (4)$$

The room was constructed by first building a framework using standard aluminium frames; thereafter, 1100 aluminium alloy panels were bolted over these frames, which served as the conductive body of the wall, floor, and ceiling. The central pole was composed of C1220T copper alloy pipes and C1100 copper alloy plates, connected via braze welding. The lumped capacitors were bolted to the structure, as shown in Supplementary Fig. 3, using printed circuit boards containing electrodes with an immersion gold surface finish. The panels composing the structure were carefully connected to ensure that high conductivity is maintained. As shown in Fig. 2(**b**), the floor of the structure was lifted by approximately 200 mm for facilitating construction and to alleviate the effect of coupling with the building floor. The test room was built using ordinary hardware and tools; special construction techniques were not required.

For the demonstration of the furnished room (Fig. 5), we covered the ceiling and walls of the room with light-weight styrene boards to provide the appearance of an actual room. In addition, the floor was covered with PVC flooring material. We placed the walls and floor materials along the charging area shown in Supplementary Fig. 2 to form a rectangular room volume. Ordinary pieces of furniture were also placed within the room to further highlight the applicability of wireless power transfer in daily environments. The room-scale resonator's quality factor and resonant frequency were not significantly affected by these furnishing materials and the people present within the room.

**FEM simulations for determining eigenmodes of the structure**

For extracting the eigenmodes of the structure, eigenfrequency analyses were conducted using the FEM-based EM field solver, COMSOL Multiphysics (RF module). Through these simulations, the resonant frequencies, current distributions, and magnetic fields of the resonant modes were obtained. As experimental values of the quality factor are typically lower than the simulated values, we used the measured quality factor for the efficiency and safety evaluations.

**Determination of efficiency based on the coupled mode theory**

The three parameters required to determine the transmission efficiency between coupled resonators are the quality factor of the transmitter, quality factor of the receiver, and the coupling coefficient between the transmitter and the receiver. For the coupled mode theory-based analysis, the following expression of the transfer efficiency between two coupled resonators with identical frequencies was used[19,25]:

$$\eta_{\max} = \frac{\chi}{\left(1 + \sqrt{1+\chi}\right)^2} \quad (5)$$

$$\chi = \frac{4Q_1 Q_2 |\kappa|^2}{\omega_0^2} \quad (6)$$

Here, $Q_1, Q_2, \kappa$, and $\omega_0$ represent the quality factor of the transmitter (room-scale resonator), quality factor of the receiver, coupling rate between the transmitter and receiver, and angular resonant frequency of the two resonators, respectively. This expression yields the transfer efficiency when the load is tuned to a value that maximises efficiency. This is reasonable considering the development of maximum efficiency point tracking methods[18,19]. Measured values of the quality factor were used because it is difficult to obtain accurate values through simulations. FEM simulations were used to extract the coupling rate $\kappa$ through the following equations[19,21]:

$$\kappa = \frac{\sqrt{2}}{4} \cdot \frac{\omega_0 \beta}{\sqrt{L_2 \alpha}} \quad (7)$$

$$\alpha = \iiint_V \frac{\mu_0}{2} \cdot |\boldsymbol{H}|^2 \mathrm{d}V \quad (8)$$

$$\beta = \iint_A \mu_0 \boldsymbol{H} \cdot \boldsymbol{n} \mathrm{d}A \quad (9)$$

Here, $\alpha$ is the total magnetic energy stored in the room, and $\beta$ is the flux intersecting the receiver. $\boldsymbol{V}$ is the entire volume of the room, $\boldsymbol{n}$ is the unit normal vector of the receiver surface, $\omega_0$ is the angular resonant frequency of the QSCR, $\boldsymbol{A}$ is the area enclosed by the receiver, and $L_2$ is the receiver inductance.

For determining efficiency based on the coupled mode theory, the centre of the receiver resonator was placed at 50 mm interval grid points within the volume of the room. The considered range of receiver positions (*i.e.* coordinates of the centre of the receiver) is as follows:

$$\max(|x|, |y|) \leq 1.4 \text{ m} \tag{10}$$

$$0.1 \text{ m} \leq |z| \leq 0.9 \text{ m} \tag{11}$$

We excluded a small region around the pole and the walls to accommodate the finite size of the receiver. As the cross-sections of the room and the receiver resonator significantly differed, we assumed that the magnetic flux penetrating the receiver resonator was uniform. The efficiency calculated under this assumption was in good agreement with the measured efficiency. The orientation of the receiver resonator was set such that it maximises the power transfer efficiency at each position (*i.e.* all the magnetic fluxes were normal to the coil).

### Determination of efficiency based on two-port measurements

The measured efficiency, $\eta_{\max}$, was extracted using S-parameters measured by a VNA using the following expression established in literature[26]:

$$\eta_{\max} = K - \sqrt{K^2 - 1} \tag{12}$$

$$K = \frac{1 - |S_{11}|^2 - |S_{22}|^2 + \left|S_{11}S_{22} - S_{12}^2\right|^2}{2|S_{12}|^2} \tag{13}$$

Similar to the expression for determining efficiency using the coupled mode theory, this expression also yields the transfer efficiency when the load is tuned to a value that maximises efficiency. For each mode, the drive coil was placed at the positions shown in Fig. 3(**f**). Moreover, the receiver resonator was placed at the series of positions represented by lines (i) and (ii) in Fig. 3(**f**), with orientations that maximised the penetrating magnetic fluxes. This results in a coil orientation perpendicular to the floor because the generated magnetic field is parallel to the floor. For these evaluations, we used the receiver coil and the drive coil shown in Supplementary Fig. 7.

**Evaluation of safety**

To evaluate tissue heating in the frequencies we used, the IEEE and FCC adopt two sets of safety guidelines based on the SAR and incident EM field intensity, respectively. Because the SAR is a physical quantity directly related to tissue heating, the SAR limit works as a fundamental limitation. However, because SAR evaluations usually require powerful computation tools and resources, it is also permitted to use the "reference levels" defined by EM field intensity instead, which are set to values unlikely to exceed the SAR limits. Note that this "reference level" is derived similarly with maximum permissible exposure (MPE) in the FCC guidelines. When the EM field intensity exceeds this reference level (or MPE), extensive assessments of the SAR limits are necessary.

Because our system exceeded the reference levels when delivering several Watts of power, we directly conducted detailed SAR evaluations, as described in the "Safety" section.

**Augmenting devices with wireless power receivers**

For demonstrating room-scale wireless power transfer in a living environment, we augmented the devices (*i.e.* a fan, an LED lightbulb, and a smartphone) using wireless power transfer receivers. We composed these receivers with a receiver resonator, a loop coil, a full-bridge rectifier, and a linear voltage regulator. Supplementary Fig. 8 shows the coil placement on these augmented devices.

**Data availability**

The data that support the findings of this study are available from the corresponding author upon reasonable request.

**Code availability**

N/A

**Acknowledgements**

This work was supported by a Grant-in-Aid for JSPS Fellows JP18J22537, JST ERATO Grant Number JPMJER1501, and JST ACT-X Grant Number JPMJAX190F. The authors thank M. J. Chabalko for the insightful discussions. The authors also thank K. Narumi, H. Ogata, and T. Ikeuchi for help in the video production.

**Author contributions**

T.S., Y.K., and A.P.S. designed the research. T.S. proposed the initial concept, conceived the theory, implemented the system, performed the experiments/analysis, and wrote the manuscript. All authors reviewed and commented on the manuscript. Y.K. and A.P.S. provided resources. Y.K. supervised the project.

**Competing interests**

The authors declare no competing interests.

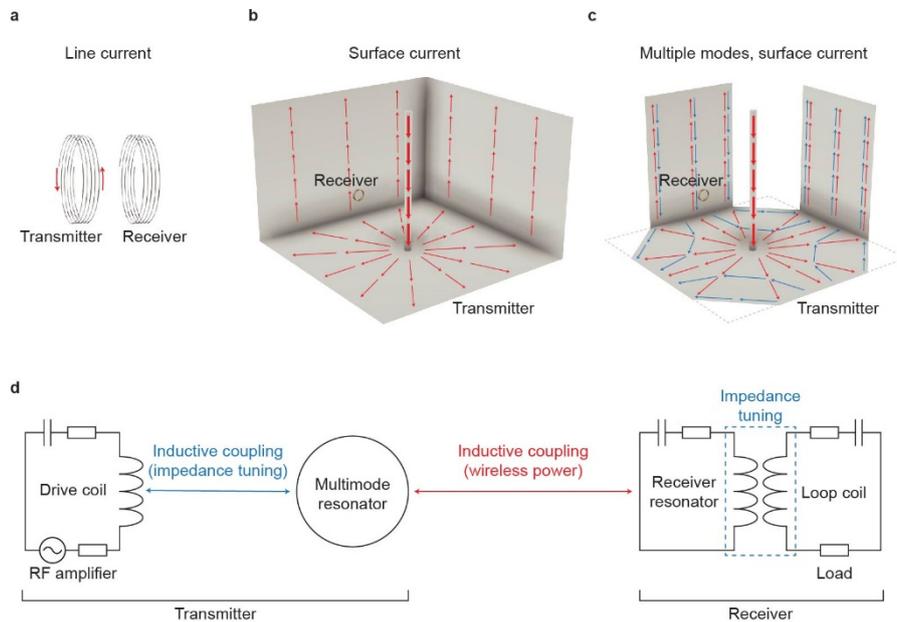

Fig. 1 Overview of multimode-quasistatic cavity resonance (M-QSCR). Conceptual diagrams of (**a**) wireless power transfer via typical coil-based resonators, which shows a narrow powering range and exhibits low efficiency in asymmetric configurations; (**b**) QSCR, which involves areas where power transfer is inefficient; and (**c**) M-QSCR, which enables efficient power delivery throughout large volumes by utilising the generated multiple magnetic field patterns. In (**b**) and (**c**), conductive sheets at the front are omitted for visibility. (**d**) Proposed wireless power transfer system based on M-QSCR; the room-wide M-QSCR was stimulated via an external drive coil (Supplementary Fig. 7(**b**)). The receiver is composed of a high-Q receiver coil and a loop coil for impedance adjustment connected to the load (see Supplementary Fig. 7(**a**)).

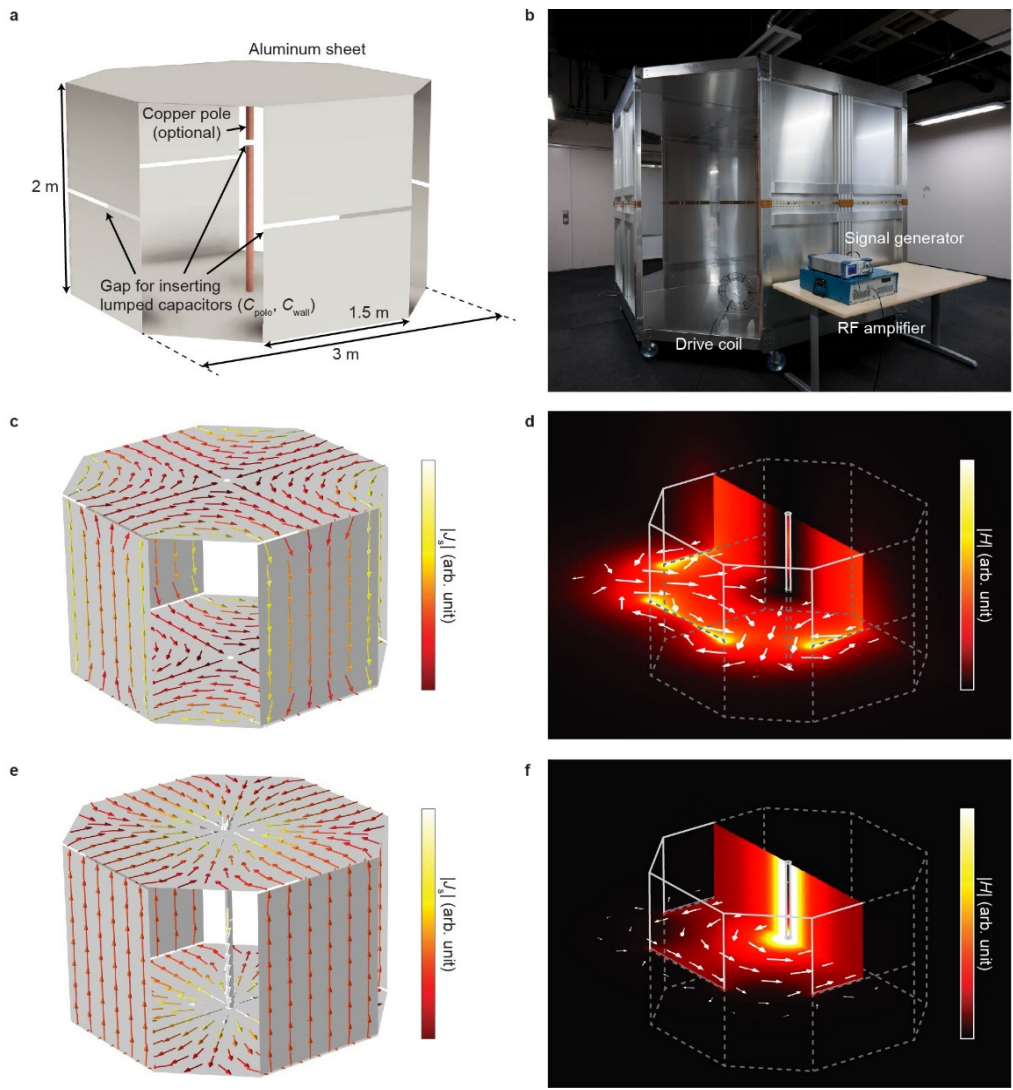

Fig. 2 Constructed room-scale resonator. (**a**) Overview and (**b**) implementation of the room-scale wireless power transfer system based on M-QSCR (see Methods for construction details). (**c**) and (**d**) show the current distribution and magnetic field distribution of the pole independent (PI) mode, respectively. (**e**) and (**f**) show the current distribution and magnetic field distribution of the pole dependent (PD) mode, respectively.

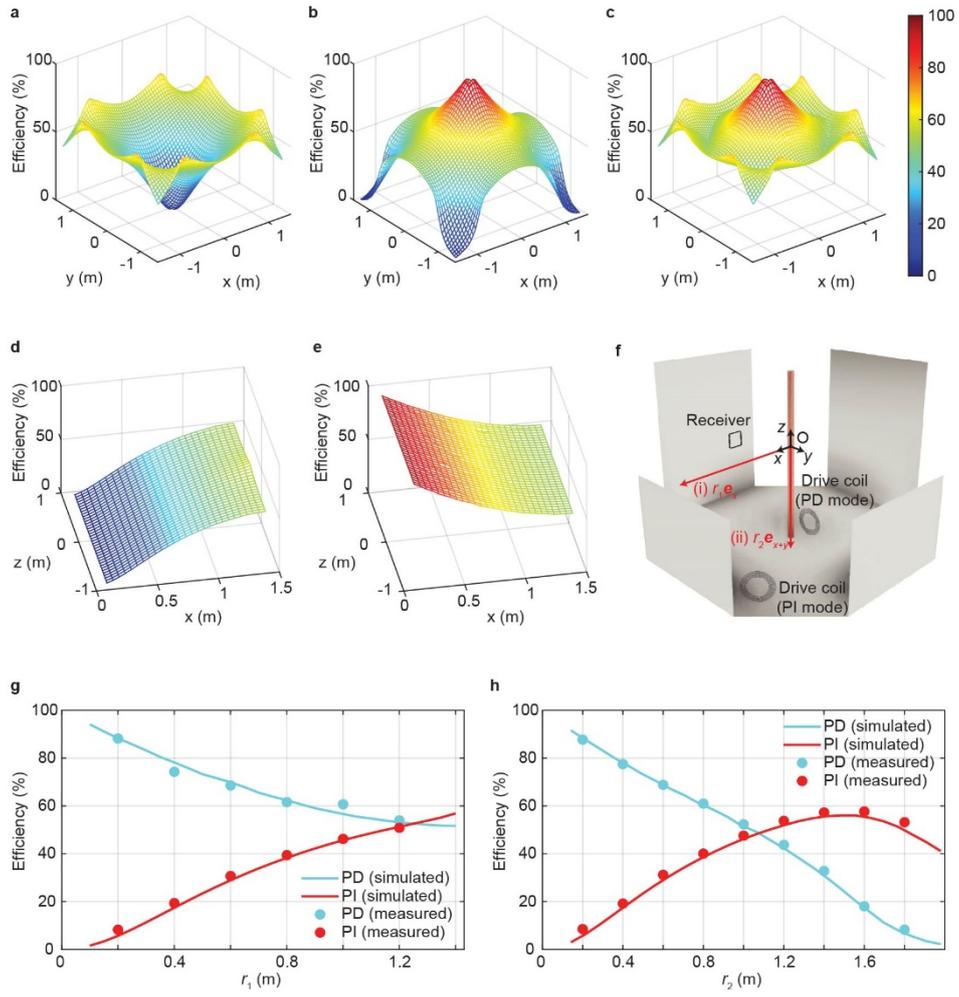

Fig. 3 Evaluation of the power transfer efficiency. Efficiencies when the receiver is placed on the $z = 0$ plane (see **f**) using the (**a**) pole independent (PI) mode, (**b**) pole dependent (PD) mode, and (**c**) both modes are shown. The power transfer efficiencies when the receiver is placed on the $y = 0$ plane using the (**d**) PI mode and (**e**) PD mode are shown. (**f**) shows the setup and coordinates used for the measurements, with conductor sheets partially omitted for visibility. The drive coil positions were altered depending on the excited resonant mode. The parameters $r_1$ and $r_2$ indicate the distance from the centre of the room. $\boldsymbol{e}_x$ and $\boldsymbol{e}_{x+y}$ are the unit vectors along the $+x$ direction and $x + y$ direction, respectively. (**g**) shows the simulated and measured power transfer efficiency when the receiver is placed on line (i) (see (**f**)). (**h**) shows the simulated and measured efficiencies when the receiver is placed on line (ii) (see (**f**)).

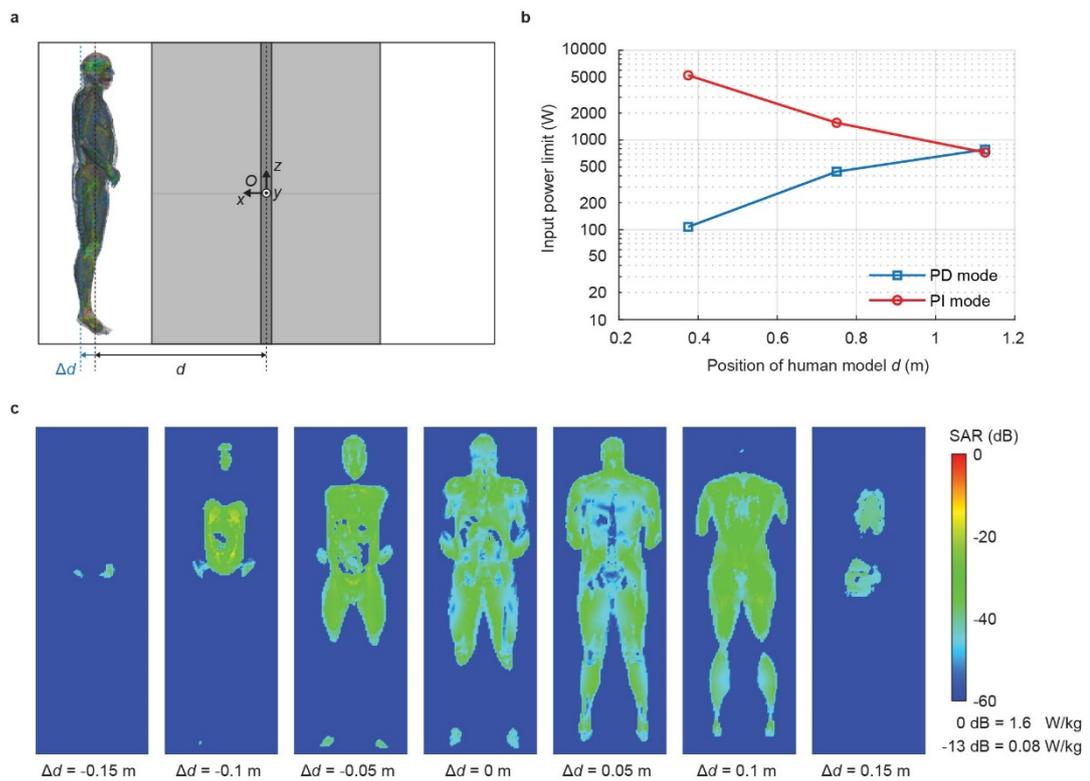

Fig. 4 Evaluation of safety based on specific absorption rate (SAR). (**a**) Side view of the anatomical human model used in SAR evaluations and its placement within the room-scale transmitter resonator. (**b**) shows the input power limit of each mode when the position of the human model was varied. (**c**) shows the SAR distribution when the input power reached the exposure limits.

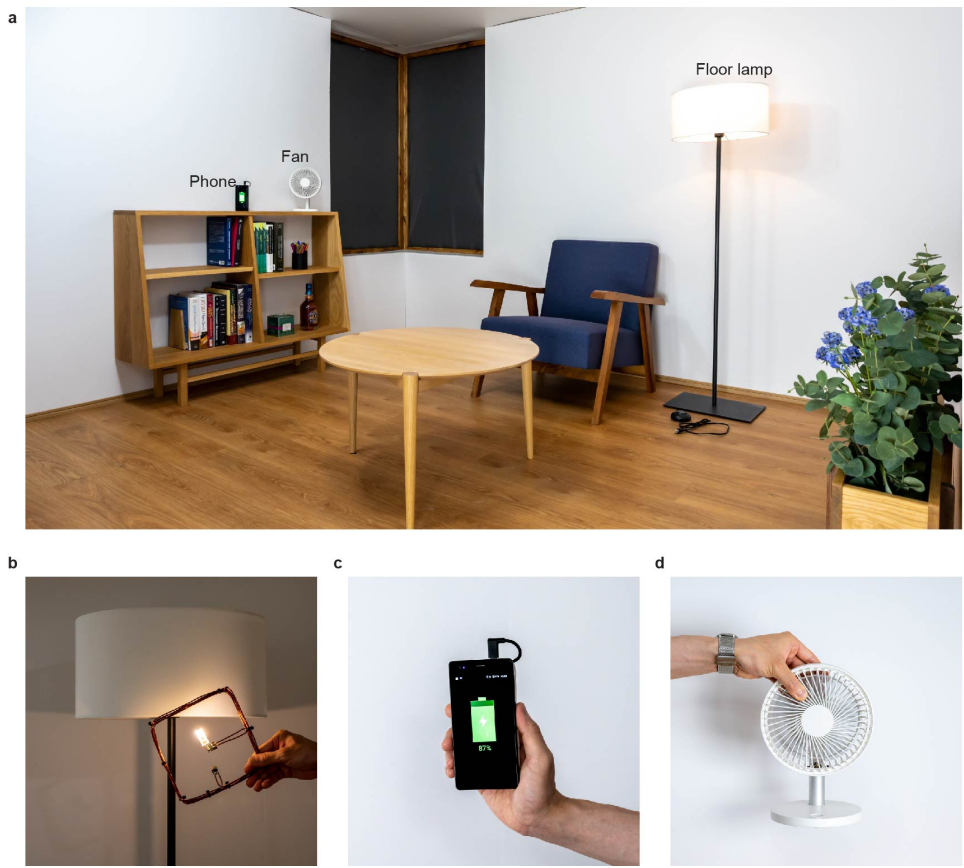

Fig. 5 Demonstration of room-scale wireless power transfer in a living environment. (**a**) shows an overview of the room; (**b**) shows a wirelessly powered light bulb (included inside the floor lamp in **a**); (**c**) shows the wireless charging of a smartphone in the air; and (**d**) shows a wirelessly powered portable fan.